\documentclass{article}
\usepackage[preprint]{neurips_2025}

\usepackage{multirow}
\usepackage{makecell}
\usepackage[utf8]{inputenc} 
\usepackage[T1]{fontenc}    
\usepackage{hyperref}       
\usepackage{url}            
\usepackage{booktabs}       
\usepackage{amsfonts}       
\usepackage{nicefrac}       
\usepackage{microtype}      
\usepackage{xcolor}         
\usepackage{amsmath}
\usepackage{amssymb}
\usepackage{mathtools}
\usepackage{amsmath}
\usepackage{amssymb}
\usepackage{graphicx}
\usepackage{listings}
\usepackage{pifont}
\newcommand{\cmark}{\textcolor{green!70!black}{\ding{51}}}%
\newcommand{\xmark}{\textcolor{red}{\ding{55}}}%

\usepackage{booktabs}
\usepackage{tabularx}

\title{FusID: Modality-Fused Semantic IDs\\for Generative Music Recommendation} 
\author{%
  {Haven Kim \quad  Yupeng Hou\quad  Julian McAuley} \\
   {University of California San Diego}
}

\begin{document}

\maketitle

\begin{abstract}

Generative recommendation systems have achieved significant advances by leveraging semantic IDs to represent items. However, existing approaches that tokenize each modality independently face two critical limitations: (1) redundancy across modalities that reduces efficiency, and (2) failure to capture inter-modal interactions that limits item representation. We introduce FusID, a modality-fused semantic ID framework that addresses these limitations through three key components: (i) \textit{multimodal fusion} that learns unified representations by jointly encoding information across modalities, (ii) \textit{representation learning} that brings frequently co-occurring item embeddings closer while maintaining distinctiveness and preventing feature redundancy, and (iii) \textit{product quantization} that converts the fused continuous embeddings into multiple discrete tokens to mitigate ID conflict. Evaluated on multimodal next-song recommendation (i.e., playlist continuation) benchmark, FusID achieves zero ID conflicts—ensuring each token sequence maps to exactly one song—and mitigates codebook underutilization, while outperforming the baselines in MRR and Recall@$k$ ($k$=1,5,10,20).
\end{abstract}

\section{Introduction}
Recommender systems are increasingly adopting generative paradigms that frame recommendation as a sequence generation task~\cite{rajput2023recommender, zheng2024adapting, deldjoo2024review, hou2025generative, hou2025actionpiece, yupeng_cikm_tutorials2025}. Central to this paradigm shift is the use of \textit{Semantic IDs} (SIDs)—discrete token sequences that encode items by mapping continuous representations—typically produced by item encoders—into discrete codebooks for autoregressive language models. In the music domain, semantic IDs successfully leverage item modalities associated with single track to extract item token sequences, which are used to design generative next-song recommendation (i.e., playlist continuation)~\cite{mei2025semantic} and conversational agents~\cite{doh2025talkplay, doh2025tools}.

Despite this promise, the implementation of SID-based generative recommenders in multimodal settings faces critical challenges. First, existing approaches~\cite{doh2025talkplay, doh2025tools} that learn separate tokenizers for each modality (e.g., audio, tag, lyrics) independently encode similar semantic information multiple times. This leads to substantial redundancy in the learned codebooks—this results in unnecessarily large vocabulary sizes that not only strain the capacity of language models, but also increase computational overhead during training and inference.  Second, by treating modalities in isolation, these methods fail to capture the rich inter-modal interactions~\cite{baltruvsaitis2018multimodal, zhao2024deep} and complementary information that naturally exist across different views of the same item; for instance, the emotional tone conveyed through audio features may align with lyrical themes in text, yet independent encoding cannot leverage such synergistic combinations~\cite{wang2025multimodalrepresentationdisentangledinformationbottleneck} to build more comprehensive and discriminative item representations. These limitations motivate the need for a unified framework that can jointly encode multimodal information while eliminating redundancy and capturing synergistic relationships across modalities.

\begin{figure*}[!t]
\centering
\includegraphics[width= \linewidth]{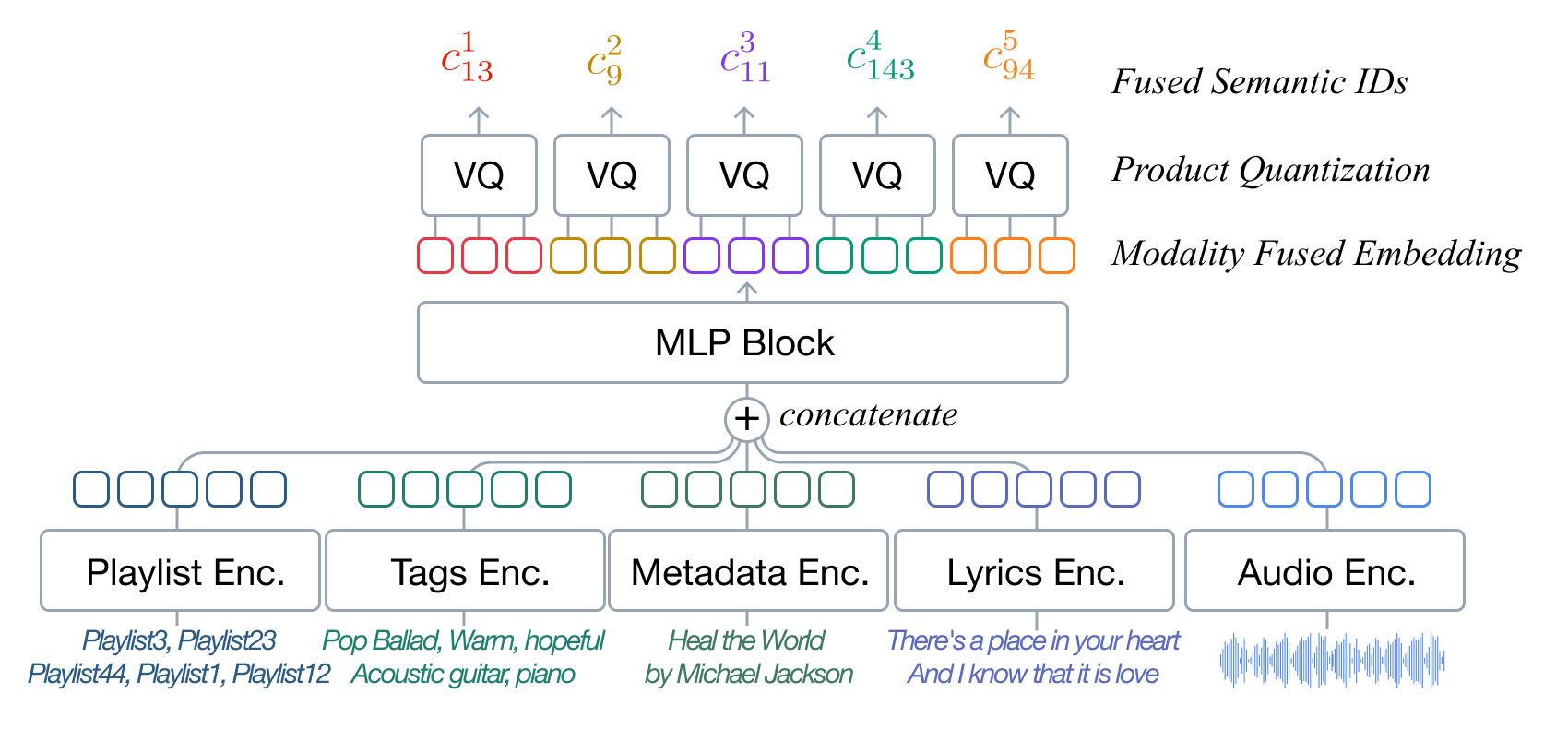}
\vspace{-3mm}
\caption{Overview of Modality-Fused Semantic IDs.}
\label{fig:main-overview}
\vspace{-3mm}
\end{figure*}

In this paper, we introduce \textbf{FusID} (Modality \textbf{Fus}ed Semantic \textbf{ID}s), a novel framework that learns modality-fused semantic IDs to address the aforementioned challenges. As illustrated in Figure~\ref{fig:main-overview}, FusID consists of three stages: \textit{(1) Multimodal Fusion Modules:} We design a simple fusion network that jointly processes multiple modalities to learn inter-modal interactions. \textit{(2) Representation Learning:} We employ a representation learning objective that brings together embeddings of frequently co-occurring items across different contexts, while pushing apart embeddings of infrequently co-occurring items. Additionally, we adopt a regularization loss that ensures all embedding elements carry important information and prevents feature overlap between tokens, maintaining distinctiveness and avoiding redundancy in the learned representations. \textit{(3) Product Quantization:} We apply product quantization to decompose the fused embeddings into multiple subspaces, generating composite semantic IDs that achieve perfect uniqueness with zero ID conflicts—guaranteeing that each token sequence maps to exactly one item.

\section{FusID: Modality-Fused Semantic IDs}

Given an item with $m$ modalities (e.g., playlist context, tags, metadata, lyrics, and audio), we first extract modality-specific features $M_1, \dots, M_m$ using pretrained encoders. FusID then employs a single fusion network that jointly consumes the concatenation of $(M_1, \dots, M_m)$ and produces a fused embedding matrix $E \in \mathbb{R}^{n \times d}$, where each row $e_i \in \mathbb{R}^{d}$ corresponds to the $i$-th semantic sub-embedding for that item. These $n$ sub-embeddings are later converted into $n$ semantic IDs via product quantization (Section~\ref{subsec:quantization}), forming the final token sequence for each item. 

\subsection{Modality Fusion Modules and Representation Learning}
To combine information across modalities, we employ a single fusion network $f_\theta$. This network takes the $m$ modal features $(M_1, \dots, M_m)$ as input, where each $M_i$ represents the $i$-th modality for an item, and outputs a fused embedding $E = f_\theta([M_1; \dots; M_m])$. We interpret $E$ as $n$ distinct embeddings $\{e_1, \dots, e_n\}$, where each $e_i \in \mathbb{R}^{d}$ is later quantized. The fusion network is trained with two types of loss functions: a contrastive loss that encourages embeddings of frequently co-occurring items to be close across contexts, while separating embeddings of items that rarely co-occur, and a regularization loss that prevents redundancy and collapse across the $n$ sub-embeddings.

\textbf{Contrastive Loss:} We employ a contrastive loss, which minimizes the energy of positive samples while maximizing that of negative samples~\cite{lecun2022path}, to encourage embeddings of frequently co-occurring items to be close while pushing apart embeddings of infrequently co-occurring items. Given a pair of items $(i, j)$ with fused embeddings $E_i$ and $E_j$, we compute the Euclidean distance $D(E_i, E_j)$ between them. Let $y \in \{0, 1\}$ indicate whether $(i, j)$ is a positive pair ($y=1$) or negative pair ($y=0$). The contrastive loss is defined as:

\vspace{-5mm}
\begin{equation}
\mathcal{L}_{\text{cont}} = \mathrm{MSE}\bigl(\lvert 1 - y \rvert,\; D(E_i, E_j)\bigr)
\end{equation}
\vspace{-5mm}

For contrastive pair mining, we use normalized co-occurrence, treating high-scoring pairs as positives and low-scoring pairs as negatives.

\textbf{Regularization Loss:} To prevent redundancy and ensure all sub-embeddings carry useful information, we apply a regularization loss $\mathcal{L}_{\text{reg}} = \mathcal{L}_{\text{cov}} + \mathcal{L}_{\text{var}}$ , inspired by VICReg~\cite{bardes2021vicreg}. The covariance loss $\mathcal{L}_{\text{cov}}$ encourages the $n$ sub-embeddings $\{e_1, \dots, e_n\}$ to capture complementary information by penalizing correlations between them. Because our goal is to avoid feature overlap within each embedding ($E_i = {e_{1}, \dots, e_{n}}$), rather than applying the element-wise penalty used in Bardes et al.~\cite{bardes2021vicreg}, we employ the block-wise cross-covariance terms between the (n) chunks (thus considering $\binom{n}{2}$ pairs):

\vspace{-4mm}
\begin{equation}
\mathcal{L}_{\mathrm{cov}} = \frac{2}{n(n-1)} \sum_{i=1}^{n} \sum_{\substack{j=1 \\ i < j}}^{n} \frac{1}{d} \bigl\| \mathrm{Cov}(e_i, e_j) \bigr\|_F^2
\end{equation}
\vspace{-4mm}

where $\mathrm{Cov}(e_i, e_j)$ is the cross-covariance matrix between sub-embeddings $e_i$ and $e_j$. 

The variance loss $\mathcal{L}_{\text{var}}$ ensures each feature dimension maintains sufficient variance and prevents the embeddings from collapsing into a trivial solution (e.g., all vectors becoming zero or constant), which would also make the covariance loss trivially zero~\cite{bardes2021vicreg}:

\vspace{-4mm}
\begin{equation}
\mathcal{L}_{\text{var}} = \frac{1}{n \times d} \sum_{c=1}^{n \times d} \max\Bigl(0,\; \gamma - \sqrt{\mathrm{Var}(e_{\cdot, c}) + \epsilon}\Bigr)
\end{equation}
\vspace{-3mm}

where $\gamma=1$ is the target standard deviation and $\epsilon$ is a small constant for numerical stability. The total training objective combines both losses: $\mathcal{L}_{\text{total}} = \mathcal{L}_{\text{cont}} + \alpha \,\mathcal{L}_{\text{reg}}$, where $\alpha$ controls the regularization strength. We empirically set $\alpha = 0.2$ in all our experiments.

\subsection{Quantization Modules}~\label{subsec:quantization}
After training the fusion network, we convert the continuous fused embeddings into discrete semantic IDs using product quantization. For each item, we take the fused embedding matrix $E \in \mathbb{R}^{n \times d}$ and view its $n$ rows as sub-embeddings $(e_1, \dots, e_n)$, each in $\mathbb{R}^{d}$ (with $d = 128$). Each sub-embedding $e_i$ is assigned to one of $K=1024$ clusters using $k$-means clustering in its corresponding subspace, using songs included in the training set. When $e_i$ is assigned to the $x_i$-th cluster ($x_i \in \{1, \dots, 1024\}$), it is tokenized as $c_{i,x_i}$.

Thus, a single song is represented as a sequence of $n$ tokens $(c_{1,x_1}, \dots, c_{n,x_n})$, where each position $i$ has its own codebook of size $1024$. This product-quantized design yields a large combinatorial semantic ID space while ensuring that each token sequence maps to exactly one item, eliminating ID conflicts and enabling scalable generative recommendation over fused multimodal representations.

\section{Experiments}

\textbf{Dataset:} Our dataset of choice is the Million Playlist Dataset (MPD)~\cite{chen2018recsys}. Although the original MPD contains 2,262,292 unique songs, we obtained complete feature embeddings (tags, metadata, lyrics, and audio) for only 537,042 songs. Therefore, we filtered out playlists containing fewer than six of these valid songs, as our downstream task requires sufficient context to generate sequence recommendations. This process yielded 926,689 playlists. On average, a playlist in this filtered dataset contains 36.81 songs. Finally, we partitioned the data into training (80\%), validation (10\%), and test (10\%) sets. 

\textbf{Evaluation:} We evaluate our approach from two perspectives: (1) semantic ID quality and (2) generative recommendation performance. For Semantic ID quality, We evaluate three key metrics: (1) \textit{Codebook Underutilization Rate (CUR)}, which measures the percentage of unused codebook entries; (2) \textit{Cardinality}, which counts the number of unique items representable by semantic IDs; and (3) \textit{Conflict Rate}, which measures the percentage of items that share the same semantic ID with other items. For generative recommendation performance, we evaluate the performance of a language model trained using these semantic IDs on the task of generative recommendation. Following a prior work~\cite{jannach2015beyond, hariri2012context}, we measure performance using MRR and Recall@$k$ (where $k=1,5,10,20$, which evaluate the models' ability to predict the next song. Specifically, for a playlist of length $l$, we use the first $l-1$ tracks as context when $l \leq 30$, and the first 30 tracks otherwise; the subsequent track in the playlist is treated as relevant ground-truth item.

\textbf{Implementation Details:} Following prior work~\cite{doh2025tools}, we extracted embeddings for four modalities: tags ($M_{\text{tag}} \in \mathbb{R}^{1024}$), metadata ($M_{\text{metadata}} \in \mathbb{R}^{1024}$), lyrics ($M_{\text{lyric}} \in \mathbb{R}^{1024}$), and audio ($M_{\text{audio}} \in \mathbb{R}^{512}$), where $M_{\text{tag}}$, $M_{\text{metadata}}$, and $M_{\text{lyric}}$ are extracted using a pre-trained text encoder~\cite{yang2025qwen3}, and $M_{\text{audio}}$ is extracted using a pre-trained audio encoder~\cite{wu2023large}. The fifth modality, playlist co-occurrence ($M_{\text{playlist}} \in \mathbb{R}^{128}$), was obtained by following the approach of prior work \cite{doh2025talkplay} where a word2vec~\cite{word2vec2013}-style embedding model learns to capture implicit relationships between songs by treating each track as a ``word'' and playlists as ``documents'', using the training set.

The input vector $M_{\text{concat}} \in \mathbb{R}^{3072}$, which concatenates five distinct vectors, is processed by a projection head to produce the final fused embedding $E \in \mathbb{R}^{n \times 128}$. This head consists of a linear layer mapping from 3072 to 2048 dimensions, followed by Batch Normalization~\cite{ioffe2015batch} and a ReLU activation. A final linear layer then projects this 2048-dimensional representation to an $n \times 128$ output, which is followed by Layer Normalization~\cite{ba2016layer}. For quantization, we use 1024 clusters for each $n$ sub dimension ($n=5$). The FusID was trained with a batch size of 128 and a learning rate of 5e-4. For the generative recommendation task, we trained a GPT-2~\cite{radford2019language} style decoder-only structure on sequences of semantic IDs and evaluated its ability to model playlist continuation by ranking the remaining tracks given the preceding context.

\section{Results}

\textbf{Multimodal Semantic IDs:} Table~\ref{tab:codebook_quality} presents the codebook quality comparison between our FusID approach and baselines. The results demonstrate that FusID with both contrastive and regularization losses achieves nearly perfect performance across all metrics. Specifically, it achieves a 0\% codebook underutilization rate (CUR) on the full dataset and a 0.02\% CUR on the test set, indicating that only a single codebook entry remains unused. In addition, it successfully encodes all 537,042 songs in the dataset (i.e., perfect cardinality), with zero conflicts. This indicates that every item receives a unique semantic ID, eliminating ambiguity in the generative recommendation process.

In contrast, the ablation study underscores the critical role of the regularization loss. When only the contrastive loss is employed (FusID Ablation), the model attains a zero codebook underutilization rate (CUR) on both the full dataset and the test set; however, it encodes 514,929 of the 537,042 songs in the full dataset and 178,661 of the 193,124 songs in the test set, with conflict rates of 6.23\% and 11.02\% for the full and test sets, respectively.

The baseline TalkPlay method shows markedly degraded performance in terms of CUR, with 0.43\% on the full dataset and 11.78\% on the test set. It also exhibits semantic ID conflicts, with conflict rates of 0.31\% and 0.27\% and cardinalities of 536,080 of 537,042 songs and 192,836 of 193,124 songs for the full and test sets, respectively.

Collectively, these results demonstrate that our proposed method effectively learns high-quality semantic IDs and improves generative recommendation performance.


\begin{table}[!t]
\centering
\caption{Semantic ID Evaluation.}
\label{tab:codebook_quality}
\begin{tabular}{lcccccccc} 
\toprule
& & & \multicolumn{2}{c}{CUR ($\downarrow$)} & \multicolumn{2}{c}{Cardinality ($\uparrow$)} & \multicolumn{2}{c}{Conflict Rate ($\downarrow$)} \\
\cmidrule(lr){4-5} \cmidrule(lr){6-7} \cmidrule(lr){8-9}
& $\mathcal{L}_{\text{cont}}$ & $\mathcal{L}_{\text{reg}}$ & All & Test & All & Test & All & Test \\
\midrule
Baseline~\cite{doh2025talkplay}          & -         & -       & 0.43  & 11.78     & 536080 & 192836  & 0.31 & 0.27 \\
FusID (Ablation)  & \cmark    & \xmark  & 0.00  & 0.00      & 514929 & 178661 & 6.23 & 11.02 \\
FusID (Ours)      & \cmark    & \cmark  & 0.00  & 0.02      & 537042 & 193124 & 0.00 & 0.00 \\
\bottomrule
\multicolumn{9}{r}{\textit{Cardinality max: 537042 (All), 193124 (Test)}}
\end{tabular}%
\end{table}

\begin{table}[]
\centering
\caption{Generative Recommendation Evaluation.}
\label{tab:evaluation}
\begin{tabular}{@{}cccccc@{}}
\toprule
\multicolumn{1}{l}{\multirow{2}{*}{}} & \multirow{2}{*}{MRR} & \multicolumn{4}{c}{Recall@k} \\ \cmidrule(l){3-6} 
\multicolumn{1}{l}{} &  & k=1 & k=5 & k=10 & k=20 \\ \midrule
SASRec~\cite{kang2018self} & 2.60 & 1.05 & 4.02 & 6.85 & 9.52 \\
TalkPlay~\cite{doh2025talkplay} & 9.02 & 6.96 & 11.61 & 13.56 & 14.79 \\
FusID (Ablation) & 7.54 & 5.90 & 9.53 & 11.06 & 12.51 \\
FusID (Ours) & \textbf{9.58} & \textbf{7.40} & \textbf{12.36} & \textbf{14.41} & \textbf{15.69} \\ \bottomrule
\end{tabular}
\end{table}

\textbf{Generative Recommendation:} Table~\ref{tab:evaluation} presents the generative recommendation performance comparison between our proposed method and the baselines. We selected two baselines: SASRec~\cite{kang2018self}, an autoregressive sequential recommendation model, and TalkPlay~\cite{doh2025talkplay}. For ablation, we also compared FusID against a variant trained without the regularization loss.

The results demonstrate that FusID consistently outperforms all baselines across all metrics. Specifically, FusID achieves performance gains over TalkPlay, the best-performing baseline, including a 6.21\% improvement in MRR and Recall@$k$ improvements of 6.32\%, 6.46\%, 6.27\%, and 6.09\%, respectively. These results validate that the high-quality semantic IDs produced by FusID's multimodal fusion approach enable more effective generative recommendation, as the language model can better capture semantic relationships between items when trained on conflict-free, well-distributed semantic ID sequences.

However, when trained without the regularization loss, FusID underperforms TalkPlay while still outperforming SASRec. Compared to the full model, this ablated variant exhibits substantial performance degradation, including a 21.29\% decrease in MRR and Recall@$k$ reductions of 20.27\%, 22.90\%, 23.25\%, and 20.27\%, respectively. These findings underscore the critical role of the regularization loss in training the FusID module.

\section{Conclusion}
In this paper, we introduced FusID, a multimodal semantic ID generation framework that addresses key limitations in existing approaches for generative recommendation. By fusing multiple modalities—including tags, metadata, lyrics, audio, and playlist co-occurrence—through a contrastive learning objective and regularization loss, FusID produces high-quality semantic IDs with zero conflicts and nearly perfect codebook utilization. Our experiments on the Million Playlist Dataset demonstrate that FusID not only achieves superior semantic ID quality but also leads to significant improvements in downstream generative recommendation performance, with consistent gains across all evaluation metrics. These results highlight the importance of multimodal fusion and conflict-free semantic representations for effective generative recommendation systems.


{\small
\bibliography{reference}
\bibliographystyle{abbrv}
}
\end{document}